\documentclass[sigconf]{acmart}
\copyrightyear{2026}
\acmYear{2026}
\setcopyright{cc}
\setcctype{by}
\acmConference[RecSys '26]{20th ACM Conference on Recommender Systems}{September 27-October 02, 2026}{Minneapolis, MN, USA}
\acmBooktitle{20th ACM Conference on Recommender Systems (RecSys '26), September 27-October 02, 2026, Minneapolis, MN, USA}
\acmDOI{10.1145/3773078.3841270}
\acmISBN{979-8-4007-2284-4/2026/09}

\begin{document}

\title{Monte Carlo Power Analysis for Small-System A/B Trials}
\subtitle{Detecting Differences in Newsletter Engagement}

\author{Michael D. Ekstrand}
\email{mdekstrand@drexel.edu}
\orcid{0000-0003-2467-0108}
\correspondingauthor
\affiliation{%
  \institution{Drexel University}
  \city{Philadelphia}
  \state{PA}
  \country{USA}
}

\author{Daniel Kluver}
\email{kluve018@umn.edu}
\orcid{0000-0002-2810-7243}
\affiliation{%
  \institution{University of Minnesota}
  \city{Minneapolis}
  \state{MN}
  \country{USA}
}

\author{Karl Higley}
\email{karl@gwendydd.io}
\orcid{0009-0002-6332-8997}
\affiliation{%
  \institution{Gwendydd Research}
  \city{Buffalo}
  \state{NY}
  \country{USA}
}

\author{Bart P. Knijnenburg}
\email{bartk@clemson.edu}
\orcid{0000-0003-1341-0669}
\affiliation{%
  \institution{Clemson University}
  \city{Clemson}
  \state{SC}
  \country{USA}
}

\renewcommand{\shortauthors}{Ekstrand et al.}

\begin{abstract}
  Good experimental practice and predicting the scientific usefulness of an experiment or experimental platform require power analysis: estimating, \textit{a priori}, how likely the experiment is to detect the intended effect if indeed it exists.
  While simple experimental designs admit well-understood power analysis methods, more sophisticated experimental designs and settings often require bespoke techniques to avoid either over- or under-estimating experimental power.
  We present the Monte Carlo method we use to estimate the power of experiments intended to increase user engagement with personalized e-mail newsletters of recommended news articles.
\end{abstract}



\maketitle

\section{Introduction}

\let\svthefootnote\thefootnote
\let\thefootnote\relax
\footnotetext{
Source code for simulation and analysis is at \url{https://doi.org/10.5281/zenodo.21381743}.
Observed data subject to a data-sharing agreement.
POPROX is at \url{https://poprox.ai}.
}
\let\thefootnote\svthefootnote

When allocating limited capacity to run studies on a set of real users, it is important to prioritize experiments that are likely to produce measurable effects. 
In classical experimental design, the standard tool for this pre-experiment assessment is \textit{power analysis}~\cite{cohenStatisticalPowerAnalysis2013}: given an experiment size and assumptions about the true size and variance of the effect to be measured, how likely is the experiment to detect the effect as statistically significant?
This likelihood is called the experiment's \textit{power}.
Power analysis examines the relationship between sample size, expected effect size, and experiment power, with the most common task being assessing how many users are needed to ensure a power of 80\% (the common power-target).

Power analysis has a closed form solution for simple experimental designs, such as detecting the difference between two conditions using two means of a normally-distributed measurement.
These closed-form solutions are implemented in power analysis tools, such as G*Power~\cite{faulGPower3Flexible2007}, and introductory research methods classes frequently teach such tools~\cite{creswell2017research}.
More sophisticated experiments with more conditions, heavily-skewed outcome variables, and/or a complex nested sampling structure (e.g. considering multiple observations per user) require simulation (Monte Carlo) designs~\cite{muthenHowUseMonte2002}.
Further, we seek not only to determine whether a particular experiment is sufficiently powered, but to assess the impact of baseline product improvements on experimental power, which requires both the parameters and the outcomes of the analysis to be interpretable and manipulable in terms of the product and subscriber experience.



To support platform design and management of POPROX News, a research-focused news recommendation and experimentation platform, we used Monte Carlo power analysis with a generative model of interaction data to estimate the power of A/B tests conducted on our platform.
In this note, we report on the simulation design and parameter estimation to provide an example and starting point for others requiring similar power analyses for A/B tests.

\section{Problem Statement}

Our core recommendation product consists of a daily e-mail newsletter of personalized news recommendations, typically consisting of 5 sections each with 3 articles. 
Earlier versions contained a single list of 10 articles, and our method applies equally to both.
Due to limitations of the email platform, as well as news content licensing, our primary form of feedback from subscribers is tracking which articles the subscriber clicks to read in full.

The typical experiment is a between-subjects A/B trial, where two groups of subscribers receive newsletters from two different recommendation pipelines, in which the key outcome is increased engagement with news articles.
While more nuanced metrics and analyses may be possible with larger user populations, we formalize newsletter engagement with two primary metrics:

\begin{itemize}
    \item The \textit{newsletter click-any probability} (CAP), the probability a user will click at least one article in a newsletter.
    \item The \textit{article click-through rate} (CTR), the probability the user will click on any given article.
\end{itemize}

We compute these metrics per-user, and compare treatments by the comparing mean per-user CAP or CTR with a $t$-test.

To assess experiment designs in this context, we require (1) an accurate generative model of data collected from our subscribers that we can use to simulate experimental observation data; (2) a parameterization of that model that can be manipulated to realize different product and subscriber activity conditions; and (3) an experiment simulator that uses this model to simulate experiments, observations of subscriber response, and analyses of these observations to ultimately simulate experiment outcomes.

\begin{table}[tb]
\caption{Posterior parameter and distribution summaries (8 chains $\times$ 2.5K samples/chain). $\hat{R}=1$ shows good mixing.}
\label{tbl:params}
{\small
\begin{tabular}{lrrrrr}
 & Mean & SD & 89\% ETI & ESS (bulk) & $\hat{R}$ \\
\midrule
$\psi$ & 0.55 & 0.02 & (0.52, 0.58) & 11,243 & 1.00 \\
$\mu$ & 0.32 & 0.05 & (0.25, 0.40) & 26,489 & 1.00 \\
$\sigma^2$ & 0.77 & 0.25 & (0.46, 1.21) & 12,016 & 1.00 \\
\bottomrule
\end{tabular}}
\end{table}

\begin{figure}[tb]
    \includegraphics[width=\columnwidth]{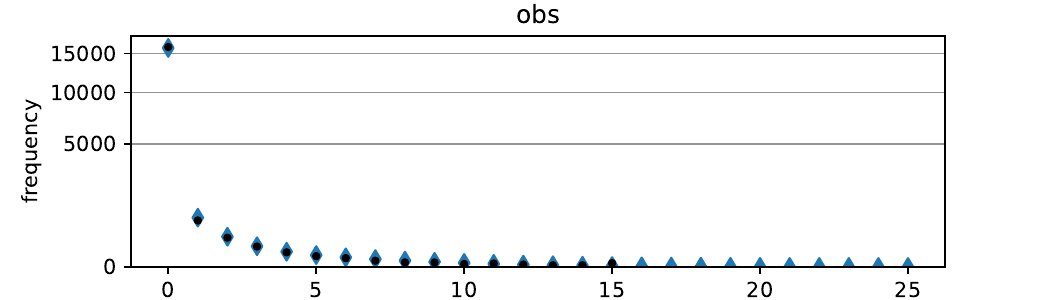}
    \Description{
    Rootogram showing that the observed distribution of click counts is within the posterior predicted distributions.
    }
    \caption{Rootogram of the fit posterior distribution (blue bands) vs. observed distribution (black dots) of $n_{\mathscr{l}}$.}
    \label{fig:posterior}
\end{figure}


\section{Generative Model of User Response}

We focus on generatively modeling $n_{\mathscr{l}}$, the number of articles clicked in newsletter $\mathscr{l}$ (sent to subscriber $u$).
From our subscriber click logs, $n_\mathscr{l}$ is a right-skewed variable with a very large number of 0s (most newsletters receive no clicks, and many subscribers never click on articles in a newsletter).
Zero-inflated count models --- either Poisson (ZIP) or negative binomial (ZINB) --- are natural fits for such data, and do not us to explicitly (and possibly arbitrarily) distinguish between active and inactive users. 
ZINB was a better fit than ZIP for our data when we ignored per-user effects\footnote{
Goodness-of-fit compared by leave-one-out expected log-posterior density (ELPD-LOO) and KL divergence between posterior and observed distributions.
} and admits a useful way to account for per-user effects that also supports our need for manipulability: since the negative binomial is Poisson with a gamma-distributed rate, we can pool Poisson rates by user.
That is, we draw a per-user Poisson rate $\lambda_u$ (``user clickiness'') from a Gamma distribution, and then model that user's click counts with a zero-inflated Poisson, yielding the following generative model:

\vskip 1ex
{\centering
$\displaystyle
\begin{aligned}\lambda_u & \sim \mathrm{Gamma}(\mu, \sigma) \\
n_{\mathscr{l}} & \sim \mathrm{Poisson}(\lambda_u) & \text{with probability $\psi$}\\
n_{\mathscr{l}} & = 0 & \text{with probability $1 - \psi$}\\
\end{aligned}$
\par}
\vskip 1ex

This model has three parameters: $\mu$ is the \textit{average user clickiness} (mean of $\lambda_u$), $\sigma^2$ is the variance of $\lambda_u$, and $\psi$ is the zero-inflation term for the final counts.
While $\psi$ and $\lambda_u$ do not map directly onto user actions, we can intuitively think of $\psi$ as modeling the probability that the user will even look at the newsletter in the first place, and $\lambda_u$ as the expected number of clicks given that they look at it. 

We used Markov-chain Monte Carlo Bayesian inference with vague priors\footnote{
We used PyMC 5.28.2 \citep{abril-plaPyMCModernComprehensive2023} on an NVidia RTX 4090 GPU for MCMC inference, and ArviZ 0.8 \citep{martinArviZModularFlexible2026} to summarize posteriors.
} to estimate $\mu$, $\sigma$, and $\psi$ from our historical subscriber click logs.
The pooled model fits our observed data better (ELPD$\uparrow$ $-4.4\times 10^3$ vs. $-6.3\times 10^3$ for un-pooled ZINB).
Fig.~\ref{fig:posterior} shows that posterior draws have a similar distribution as observed click counts $n_{\mathscr{l}}$, and Tbl.~\ref{tbl:params} summarizes parameter posteriors\footnote{
An alternate model with per-user $\psi_u$ terms mixed poorly ($\hat{R}>=1.02$) and yielded parameter estimates with unusably high variance.
}.

With our model, we can simulate experiments that increase user click propensity by adjusting $\mu$: to model an increase of 1.5x, we can run the generative model twice with $\mu_A$ from Tbl.~\ref{tbl:params} and $\mu_B= 1.5\mu_A$, and compute CAP and CTR from the simulated $n_{\mathscr{l}}$ values.

Unfortunately the manipulable parameters of the pooled ZINB, especially $\mu$, are not in observable units or commonly-used effect size scales for two reasons: $\mu$ encodes the average click count ($\operatorname{E}[n_{\mathscr{l}}]$), not newsletter CAP ($\operatorname{P}[n_{\mathscr{l}}>0]$); and the quantities are also affected by $\psi$.
Standard effect sizes are in terms of the log-odds of the observed quantity (e.g., CAP).
We used SymPy to derive the relationship between $\mu$ and CAP and solve for $\mu_B/\mu_A$ given base model parameters and a target CAP odds ratio.
With our simulation's current parameters, $\mu_B/\mu_A = 1.34$ produces an observed CAP odds ratio of 1.5 (boundary between small and medium effect~\cite{cohenStatisticalPowerAnalysis2013,chenHowBigBig2010}).

\begin{figure}[bt]
    \includegraphics[width=\columnwidth]{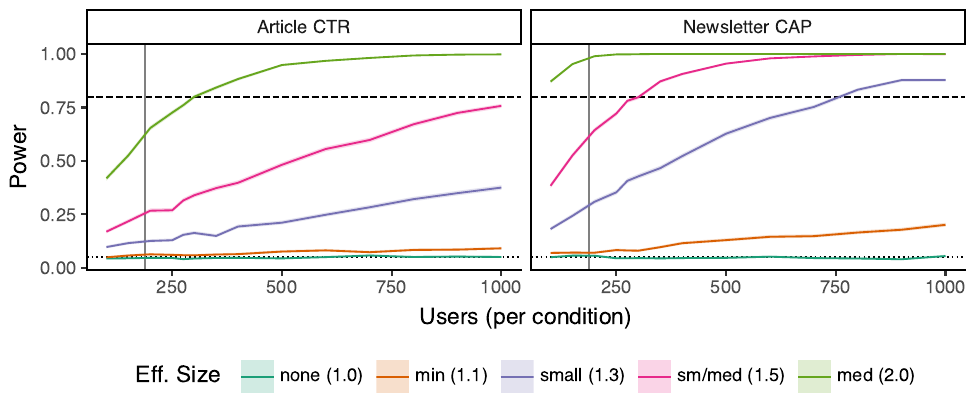}
    \Description{Power curves for both article CTR and newsletter CAP, showing that the current platform is not yet large enough to reliably detect effects of size 1.5, but can detect medium newsletter CAP changes of size 2.0.}
    \caption{Power of detecting various effect sizes (expressed as CAP odds ratios) with $t$-tests over observable metrics.  Vertical lines indicate the our current user count. Bands are 95\% CIs.}
    \label{fig:power-ttest}
\end{figure}

\section{Example Results}

Figure~\ref{fig:power-ttest} shows power curves output by our simulator (2000 iterations per condition) for varying experiment sizes, assuming fixed exposure (newsletters / user).
The simulator design allows us to programmatically compare arbitrary hypothetical experimental conditions. 
We used binary search to find the required users to detect an observable increase in CAP: a 1.5 odds ratio effect requires 300--400 users per condition, while 1.3 requires 750--800 users.

\section{Conclusion}
The key take-away of this paper is that establishing a reasonable and controllable model of user behavior is a necessary precondition for conducting accurate and informative power analyses for A/B experiments. 
In our case, designing a model that fit well (both computationally and mathematically) and provides suitable control for simulating various experimental conditions was nontrivial, but the resulting model has proved critically useful for us to get first-pass estimates of the behavior and power of statistical tests for our online recommender experiments.
Future work includes extending the model to within-subjects designs and validating against actual experimental results as more experiments are run on the platform.

\begin{acks}
This work was supported by the U.S. National Science Foundation under grants 24-09199, 22-32551, 22-32552, and 22-32690.
\end{acks}



\bibliographystyle{ACM-Reference-Format}
\bibliography{power-analysis}

\end{document}